\documentclass[runningheads]{llncs}
\usepackage[misc]{ifsym} 
\usepackage{graphicx}
\usepackage{amsmath}
\usepackage{booktabs}
\usepackage{siunitx}
\usepackage{multirow}
\usepackage{subfigure}
\usepackage{url}
\begin{document}
\title{Contrastive Rendering for Ultrasound Image Segmentation}

\author{Haoming Li\inst{1,2}\thanks{Equal contribution. \textrm{\Letter} Corresponding author: \email{nidong@szu.edu.cn}}, Xin Yang\inst{1,2\star}\and Jiamin Liang\inst{1,2}\and Wenlong Shi\inst{1,2} \and Chaoyu Chen\inst{1,2} \and Haoran Dou\inst{1,2}\and Rui Li\inst{1,2}\and Rui Gao\inst{1,2}\and Guangquan Zhou\inst{3} \and
Jinghui Fang\inst{4}\and Xiaowen Liang\inst{4} \and
Ruobing Huang\inst{1,2} \and
Alejandro Frangi\inst{1,5,6} \and \\
Zhiyi Chen\inst{4} \and Dong Ni\inst{1,2}$^{(\textrm{\Letter})}$}


\authorrunning{Li et al.}
\institute{School of Biomedical Engineering, Health Center, Shenzhen University, China\\
\and Medical UltraSound Image Computing (MUSIC) Lab, Shenzhen University, China\\
\and School of Biological Sciences and Medical Engineering, Southeast University, China
\and Department of Ultrasound Medicine, Third Affiliated Hospital of Guangzhou Medical University, Guangzhou, China
\and Centre for Computational Imaging and Simulation Technologies in Biomedicine (CISTIB), School of Computing, University of Leeds, Leeds, UK
\and  Medical Imaging Research Center (MIRC), University Hospital Gasthuisberg, Electrical Engineering Department, KU Leuven, Leuven, Belgium
}

\maketitle 

\begin{abstract}
	Ultrasound (US) image segmentation embraced its significant improvement in deep learning era. However, the lack of sharp boundaries in US images still remains an inherent challenge for segmentation. Previous methods often resort to global context, multi-scale cues or auxiliary guidance to estimate the boundaries. It is hard for these methods to approach pixel-level learning for fine-grained boundary generating. In this paper, we propose a novel and effective framework to improve boundary estimation in US images. Our work has three highlights. First, we propose to formulate the boundary estimation as a rendering task, which can recognize ambiguous points (pixels/voxels) and calibrate the boundary prediction via enriched feature representation learning. Second, we introduce point-wise contrastive learning to enhance the similarity of points from the same class and contrastively decrease the similarity of points from different classes. Boundary ambiguities are therefore further addressed. Third, both rendering and contrastive learning tasks contribute to consistent improvement while reducing network parameters. As a proof-of-concept, we performed validation experiments on a challenging dataset of 86 ovarian US volumes. Results show that our proposed method outperforms state-of-the-art methods and has the potential to be used in clinical practice.
	
\end{abstract}

\section{Introduction}
\begin{figure}[htbp]
	\centering
	\includegraphics[width=\textwidth]{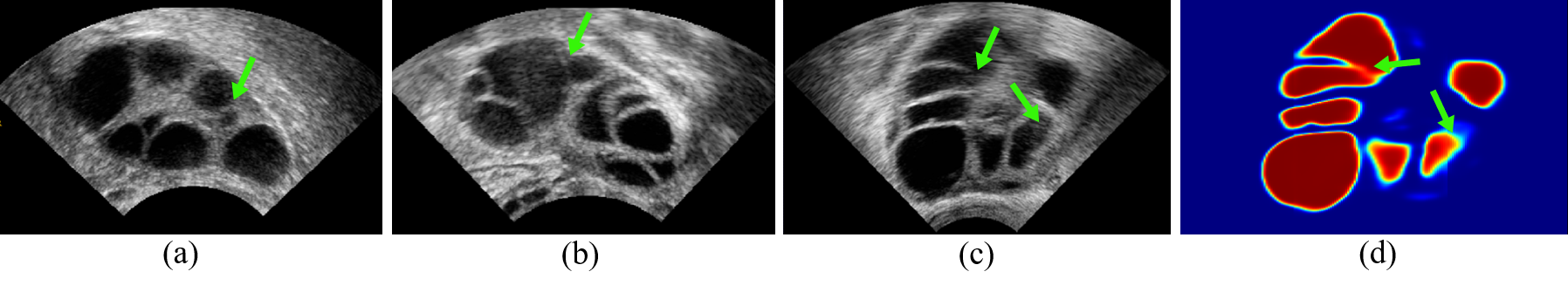}
	\caption{Illustration for (a)(b)(c) ultrasound slices of ovary and follicle. Blurry boundary and scale-varying follicles are main challenges for segmentation (green arrows). (d) Follicle probability map of (c) generated by a U-net. Underestimation and touching boundary can be observed in the map (green arrow).}
	\label{fig:challenge}
\end{figure}

Ultrasound (US) is widely accepted in clinic for routine diagnosis. Automatically segmenting anatomical structures from US images is highly desired. US image segmentation witnessed its significant improvement in this deep learning era \cite{liu2019deep}. However, due to the low contrast, low resolution and speckle noise in US images, the accuracy of these methods is hampered in ambiguous boundary regions, such as the blurred boundary or shadow-occluded parts \cite{li2019cr}. Taking the 3D ovarian US volume as an example, as shown in Fig. \ref{fig:challenge}, the adverse effects of boundary ambiguity on segmentation are obvious. The boundary among follicles are blurring due to the speckle noise. Since the follicles have irregular shapes, varying volumes and complex connection status, boundary thickness of follicles are often inconsistent (Fig. \ref{fig:challenge}(a-c)). In addition, no distinct boundary can be identified between the ovary and background tissues. All these factors can degrade the deep segmentation model with fuzzy and wrong predictions (Fig. \ref{fig:challenge}(d)). \par

In recent years, extensive attempts have been made to resolve the problems. In \cite{tu2009auto}, Tu et al. proposed a classic cascaded framework, Auto-context, to revisit the context cues in probability maps for enhancement. Local semantic features were collected by RNN in \cite{yang2018towards} to refine 2D ovarian ultrasound segmentation. Making use of auxiliary guidance to refine boundary was also studied. In \cite{chen2016dcan,zhu2019boundary}, edge-weighted mechanisms were proposed to guide the deep models to pay more attention to the edge of objects. Varying scales of object often make the boundary hard to be captured, especially on the small objects under limited imaging resolution, like the follicles in Fig. \ref{fig:challenge}. Multi-scale architectures to fuse the context information hierarchically, like Atrous Spatial Pyramid Pooling (ASPP) \cite{chen2017deeplab} are explored in medical image segmentation \cite{wang2019deep}. Although effective, these methods extract global/local semantic information for boundary refinement, but cannot approach point-level learning for fine-grained boundary identification. Recently, point-wise uncertainty estimation has been investigated to enhance image segmentation by selecting informative points for loss calculation \cite{yu2019uncertainty} and composing contour constraints \cite{wang2019boundary}. These methods selected out the ambiguous points but ignored their feature enhancement to significantly modify the predictions. \par

In this paper, to address the aforementioned problems, we propose a Contrastive Rendering (\textit{C-Rend}) framework to improve the boundary estimation in US images. By focusing on point-level boundary ambiguity and reducing it with fine-grained features, C-Rend is a general strategy with three highlights. \textit{First}, inspired by recent work \cite{kirillov2019pointrend}, C-Rend formulates boundary estimation as a rendering task. C-Rend can adaptively recognize ambiguous points and re-predict their boundary predictions via coarse and fine-grained feature enriched representation learning. \textit{Second}, to further distill discriminative features for better determination of boundary location, we selectively introduce the point-wise contrastive learning \cite{chen2020simple} to maximize the divergence among different classes. Specifically, C-Rend encourages the similarity of ambiguous points from the same class and contrastively decrease the similarity of those from different classes. Boundary ambiguities are therefore further addressed. \textit{Third}, different from previous methods which sacrifice computation overhead for performance improvement, both rendering and contrastive learning in C-Rend contribute consistent improvement while reducing the network parameters. As a proof-of-concept, we performed extensive validation experiments on a challenging dataset containing 86 3D ovarian US volumes. Results show that our proposed method outperformed strong contender and reported best results with Dice of 87.78\% and 83.49\% for the challenging problem of ovary and follicles, respectively. \par

\section{Methodology}
\begin{figure}
    \includegraphics[width=\textwidth]{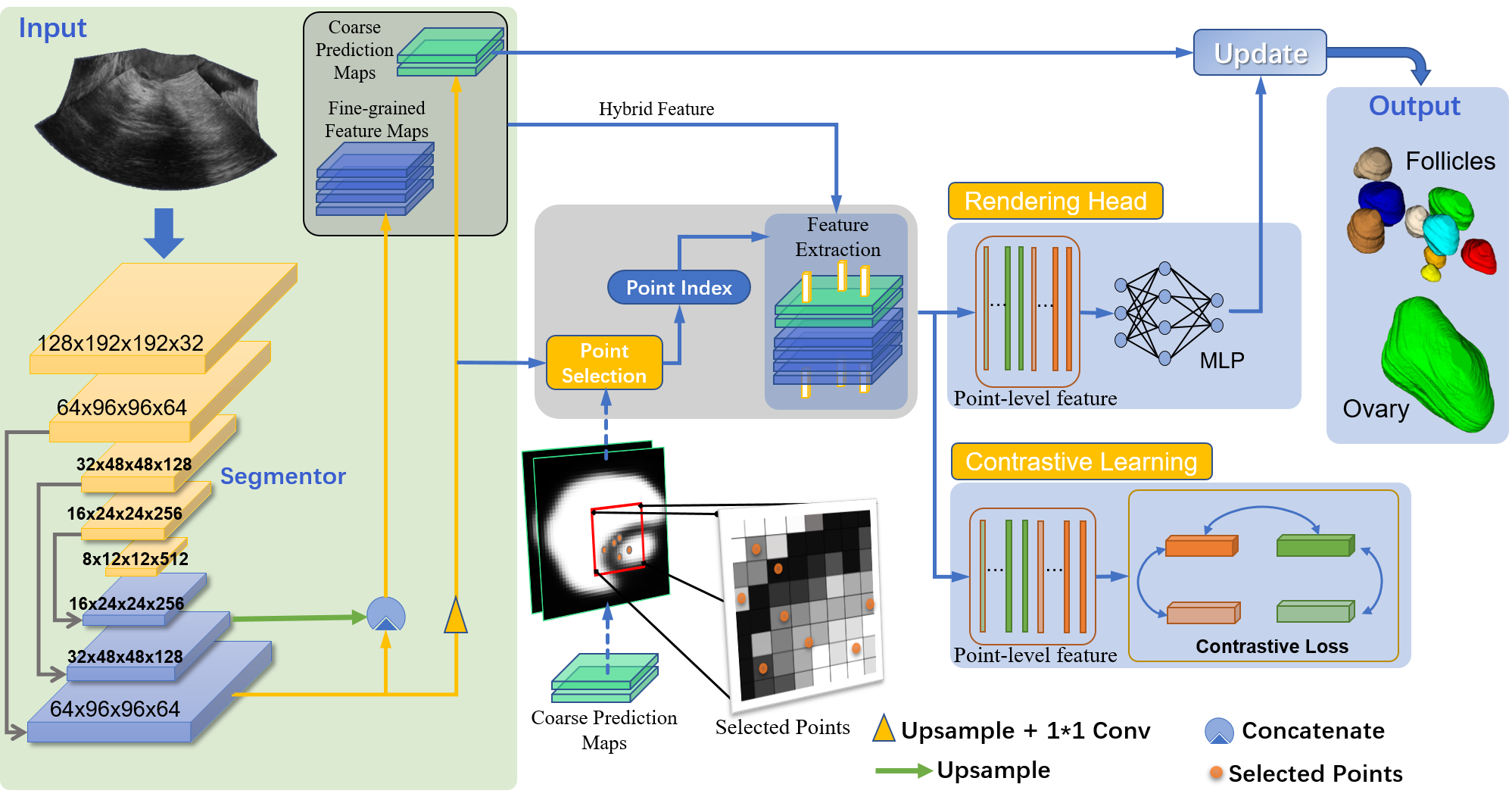}
    \caption{The overflow of our proposed contrastive rendering (C-Rend) framework.}
    \label{fig:framework}
 \end{figure}

Fig. \ref{fig:framework} illustrates the flowchart of our proposed C-Rend for 3D ovarian US volume segmentation. It mainly consists of Four main components: (1) a segmentation architecture containing an asymmetric encoder-decoder structure as the backbone, (2) point selection module to select ambiguous points that need to be re-predicted, (3) a rendering head to re-predict the label of selected points based on the hybrid point-level features, (4) a contrastive learning head to further enhance the confidence on boundary. Both rendering head and contrastive learning head contribute to the update of final segmentation output. \par
\subsection{Segmentation Architecture for Hybrid Features}
As shown in Fig. \ref{fig:framework}, we build a strong segmentation backbone based on the classic 3D U-net architecture \cite{ronneberger2015u}. The encoder contains 4 consecutive pooling layers. Between each two pooling layers, there are two convolutional layers with dilated convolution operators. While the decoder has only 3 deconvolution layers. We discard the last deconvolution layer of the original symmetric U-net to save the computation cost and reduce model parameters. Instead, we generate two kinds of predictions from lower levels of the segmentor. The first one is the coarse probability maps (full size) for the final segmentation. The second one are the fine-grained feature maps (halved size) generated from the last two layers of our segmentor with abundant channels. These two features will jointly drive our C-Rend to recognize and render the ambiguous boundary points. \par

\subsection{Rendering Head for Point-wise Re-prediction}
\subsubsection{Point Selection Strategy.}
The core of rendering head is to collect point-level hybrid feature representation to re-predict ambiguous boundaries. Whereas, directly handling all the points in the entire feature map is not only computationally expensive, but also unnecessary. Most of the boundary prediction from the segmentor present high confidence. Therefore, as shown in Fig. \ref{fig:framework}, our C-Rend firstly adopts a strategy to select $N$ points from the coarse probability map to train on \cite{kirillov2019pointrend}. Specifically, the selection strategy contains three steps. First, C-Rend randomly scatters $kN$ seeds in the coarse prediction map. Then, based on the coarse probability maps, C-Rend randomly picks out the most $\beta N$ uncertain points (e.g., those with probabilities closest to 0.5). Finally, C-Rend obtains the rest $(1-\beta)N$ points by uniformly sampling the coarse maps. During training, the selected point set is assumed to contain more points with high uncertainty than those with high confidence. In this work, we set $K=2$, $N$ equals to the volume size, $\beta=0.7$. This configuration is desired for rendering head to learn from positive and negative samples. \par

\begin{figure}
    \centering
    \subfigure[Rendering Head]{
    \includegraphics[width=0.53\textwidth]{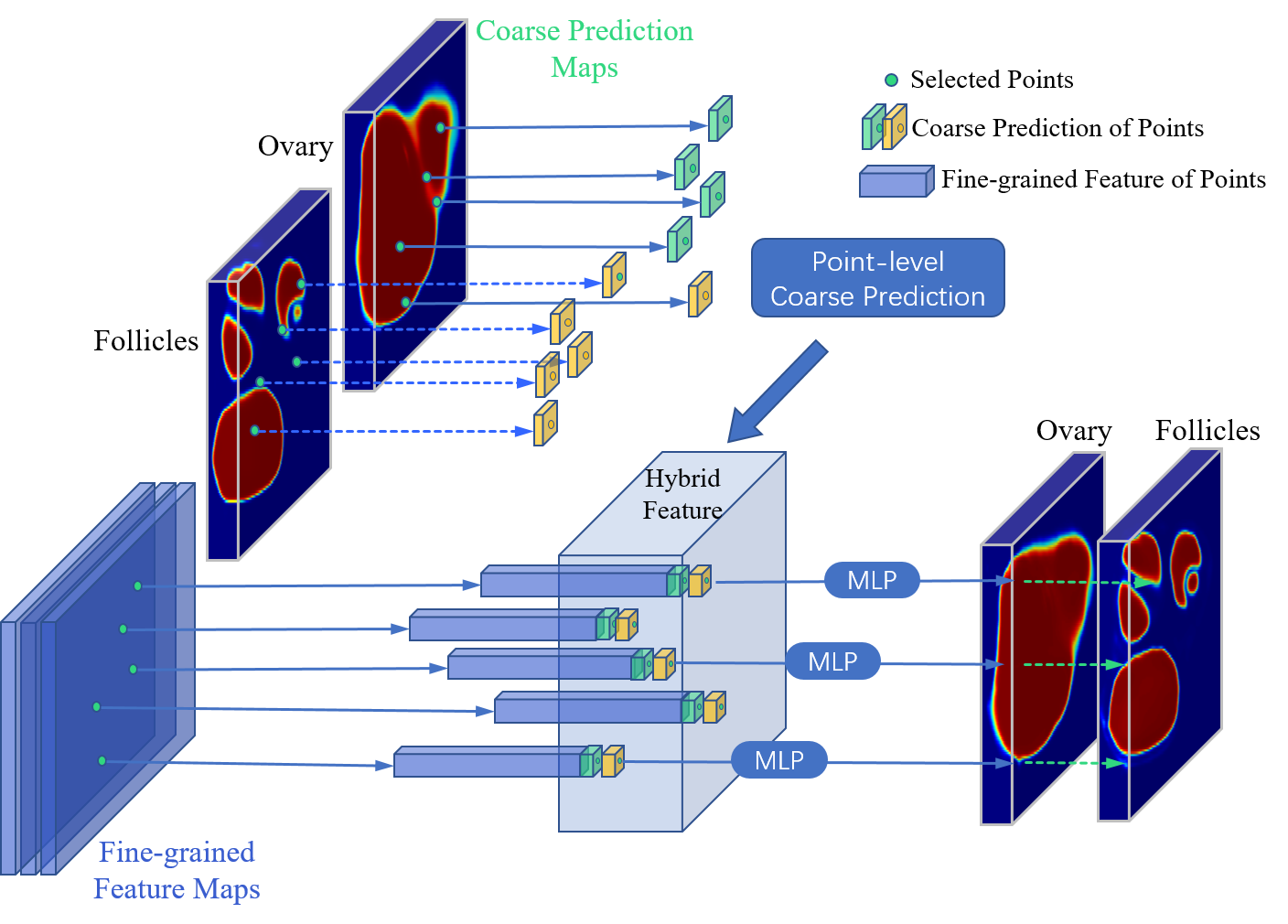}
    \label{fig:rend_head}
}
    \subfigure[Contrastive Learning]{ 
    \includegraphics[width=0.425\textwidth]{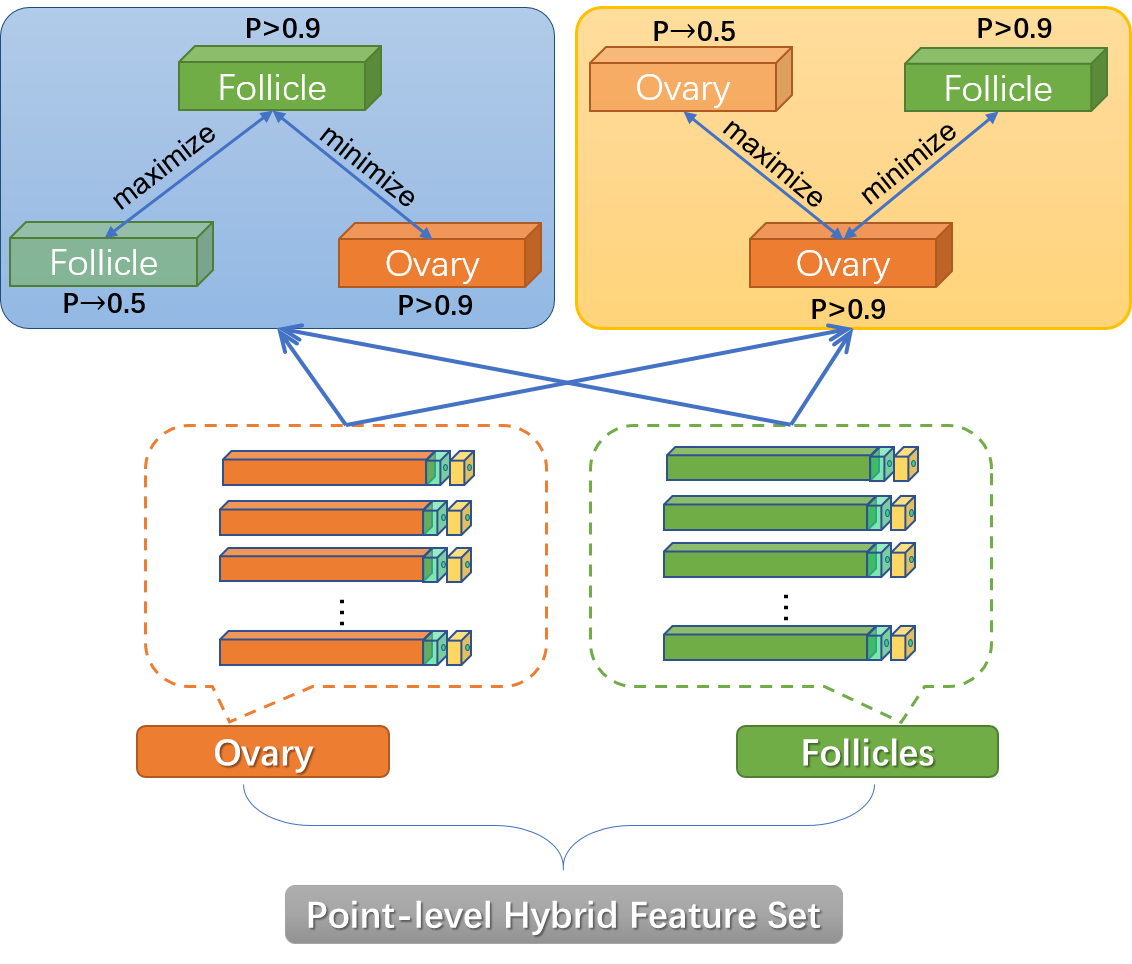}
    \label{fig:contrast_learning}
}
    \caption{(a) Details of rendering head about hybrid point-level feature and boundary re-prediction, (b) contrastive learning on the point-level hybrid representations.}
\end{figure}

\subsubsection{Rendering Head.}
Fig. \ref{fig:rend_head} elaborates the details of rendering head. With the point index provided by the selection strategy, features from the coarse probability maps and fine-grained feature maps of these points are integrated into a hybrid form. In this paper, for each selected point, a 2-dimensional probability vector is extracted from the prediction map and a 96-dimensional feature vector is obtained from the fine-grained feature maps. C-Rend concatenates these two feature vectors, and sends it to the rendering head. The mission of rendering head is to re-predict the labels of selected points. Here, we use a lightweight multi-layer perceptron (MLP) with one hidden layer as the workhorse of our rendering head. All the selected points share the same MLP (Fig. \ref{fig:rend_head}). Rendering head outputs class label for each point, therefore its training follows the standard task-speciﬁc segmentation losses. We sum the cross-entropy loss for the rendering head over all the selected points, and define it as Rendering Loss:
\begin{equation}
Loss_R =\begin{matrix} -\sum_{i=1}^N[{y_i}log(\hat{y_i})+(1-y_i)log(1-\hat{y_i})] \end{matrix}
\end{equation}
where $\hat{y_i}$ and ${y_i}$ represent the re-prediction and ground truth, respectively. $N$ denotes the number of selected points for each input volume.

\subsection{Contrastive Learning}
Considering the close connection among the boundary of objects belonging to different semantic class (in this paper, ovary and follicles), we adopt the point-wise contrastive loss to enhance the confidence on the selected ambiguous boundary points. Our contrastive loss roots in the concept of contrastive learning \cite{chen2020simple}. The basic idea is maximizing the divergence between feature vectors of two different classes, while minimizing those of the same class. Specifically, as shown in Fig. \ref{fig:contrast_learning}, based on the extracted hybrid feature vectors and probability of selected points, we enforce a doubled contrastive loss on the feature vectors from ovary and follicles. Points belonging to the follicle region with high predicted probability ($>$0.9) should have low similarity with those from ovary region, but should be as close as possible to the follicle points with high uncertainty (probability closest to 0.5). The case also applies to the uncertain points of ovary (Fig. \ref{fig:contrast_learning} top right). We use the cosine similarity to measurement of divergence. Here we build a contrastive learning loss by maximizing the disagreement among feature vectors from different categories, meanwhile minimizing the divergence among those belonging to the same category but with different confidences. The formulation of this loss, namely $Loss_C$, is:
\begin{align}
\label{eq:sim}
&Sim(\Vec{p},\Vec{q}) =\cfrac{\Vec{p}\cdot \Vec{q}}{\lVert\Vec{p}\rVert \cdot\lVert\Vec{q}\rVert}\\
\label{eq:single}
&pos_i = \sum_{k=1}^{N}{exp(Sim(p^h,p^l))}, neg_i =\sum_{k=1}^{N}{exp(Sim(p^h,q^h))}\\
\label{eq:total_loss}
&Loss_C = \gamma - \frac{1}{N}\sum_{i=1}^{N}-log(\cfrac{pos_i}{neg_i})
\end{align}

$Sim(\Vec{u},\Vec{v})$ denotes the cosine similarity between two vectors $\Vec{u}$ and $\Vec{v}$. $N$ denotes the number of selected points. Where $p^h$,$p^l$ represent the two feature vectors in same category with high and low confidence, respectively. $q^h$ also denotes the vectors with high confidence but belonging to the other category. $\gamma$ is a constant term and set as 2.0. The overall loss function of our network is:
\begin{equation}
    L_{total} =   Loss_{ori} +\lambda_1Loss_R + \lambda_2Loss_C,
\end{equation}
where $Loss_{ori}$ denotes the main cross entropy loss on the original coarse prediction of decoder. $\lambda_1=0.8$ and $\lambda_2=0.2$ denote the weight for rendering loss and contrastive loss respectively.

\section{Experimental Results}
\subsection{Dataset and Implementation Details}
We validated our framework on the 3D ovarian US segmentation. Approved by local IRB, a total of 86 volumes were collected. The maximum original size in the dataset is 400$\times$400$\times$700. Voxel spacing is re-sampled as 0.2$\times$0.2$\times$0.2$mm^{3}$. All the ovaries and follicles were manually annotated by sonographers under the quality control of an experienced expert. All the segmentation models compared in this paper were trained on 68 volumes, the remaining 18 volumes for testing. The training dataset is further augmented with $\pm$30$^{\circ}$ rotation in transverse section. Limited by GPU memory, we resized the first and second dimension of the volume to 192$\times$192, and proportionally scaled the third dimension. The segmentor finally processes the patches with size of 192$\times$192$\times$128. 50\% overlap is set during testing. In this study, all the methods were implemented in PyTorch, in which our models ran 150 epochs in a single Titan X GPU with 12GB RAM (Nvidia Corp, Santa Clara, CA). Adam with a batch size of 1 and a fixed learning rate of 0.0001 was adopted to optimize the models. \textit{Full code will be released soon.}

\subsection{Quantitative and Quanlitative Evaluation}
We adopt a set of evaluation metrics to comprehensively evaluate the segmentation results. Among them, Dice Similarity Coefficient (DSC-$\%$), Jaccard Coefficient (JC-$\%$), Hausdorff Distance (HD-mm) and Average Surface Distance (ASD-mm) were used to evaluate the segmentation on the ovary and follicles. In addition, False Detection (FD-$\%$) and Missed Detection (MD-$\%$) were utilized to assess the follicle detection. FD and MD are defined on the segmentation and IoU (Intersection over Union). A false detection (FD) of follicle means that, the segmentation hits no follicle ground truth over 30\% IoU. A missed detection means that, each follicle in ground truth hits no segmentation over 30\% IoU. Since the number of follicle is an important index for the ovary growth staging, the error on follicle counting was also adopted. Notably, for reproducibility, we reported two-fold cross validation results for all the compared methods in this paper, and in total 36 volumes were involved for each method.

Table \ref{eval_ave3} presents the quantitative comparison among our C-Rend and four strong re-implemented methods, including a two-stage Auto-Context \cite{tu2009auto} (\textit{Context}), auxiliary edge assisted segmentation (\textit{Edge}), multi-scale fusion network (\textit{Deeplab}) \cite{chen2017deeplab}, entropy around edge (\textit{Entropy}) \cite{wang2019boundary}. It is obvious to see that, C-Rend brings about consistent improvement over the baseline U-net (1.5\% DSC on follicle, 1.2\% percent DSC on ovary), while it is hard for other competitors to improve the both. C-Rend also outperforms other competitors in both follicle and ovary segmentation on almost all the metrics. Previous global contexture, edge guidance and multi-scale based designs are hence facing difficulties in handling the ambiguous boundary in fine scales. We also conduct ablation study about the modules in C-Rend. Only with the rendering head, \textit{Rend} can already refine all the metrics over the baseline. With the constraints from contrastive learning, C-Rend gets further improvements over \textit{Rend} with a large margin (about 0.7 percent in DSC for follicle and ovary). It's worth noting that, compared to the U-net baseline, our C-Rend refines the results but does not require extra computation cost (23.55M vs. 23.56M).

\begin{table}
	\centering
	\scriptsize

	\caption{Quantitative comparisons on performance and model complexity (superscript for standard deviation)}
	\begin{tabular}{c|cccc|cccc|c}
		\toprule
		\multirow{2}*{\textbf{Method}} & \multicolumn{4}{c|}{\textbf{Follicles}} & \multicolumn{4}{c|}{\textbf{Ovary}} &\multirow{2}*{\textbf{Params}}\\
		\cline{2-9}
		& DSC & JC & HD & ASD & DSC & JC & HD & ASD \\
		\hline
		U-Net\cite{ronneberger2015u} & 82.00$^{8.8}$ &70.31$^{12.4}$ & 12.10$^{3.5}$ & 4.40$^{1.0}$ &86.57$^{9.1}$ & 76.89$^{8.4}$ & 6.96$^{2.5}$ & 1.43$^{0.9}$ & 23.56M\\
		Context\cite{tu2009auto} &81.51$^{9.5}$ &69.09$^{12.5}$ &11.82$^{2.6}$ &4.24$^{1.2}$ &87.18$^{6.3}$ &77.78$^{9.4}$ &6.89$^{2.4}$ &1.36$^{0.4}$ &47.12M\\	
		Edge\cite{chen2016dcan} &82.70$^{9.5}$ &71.18$^{10.6}$ &11.79$^{2.9}$ &4.31$^{1.1}$ &86.64$^{8.7}$ &77.08$^{7.6}$ &7.08$^{3.1}$ &1.35$^{0.7}$ &23.59M \\
		Deeplab\cite{chen2017deeplab} &81.88$^{12.3}$ &70.34$^{14.1}$ &11.94$^{3.3}$ &4.36$^{2.1}$ &86.42$^{10.2}$ &76.63$^{10.4}$ &7.04$^{2.1}$ &1.33$^{0.3}$ &25.58M\\		
		Entropy\cite{wang2019boundary} &81.64$^{7.2}$ &69.13$^{10.6}$ &11.96$^{2.78}$ &4.24$^{1.17}$ &86.76$^{6.0}$ &77.22$^{9.0}$ &7.05$^{2.2}$ &1.35$^{0.43}$ &23.56M\\
		\hline
		\textbf{Rend} &82.73$^{7.6}$ &70.95$^{8.6}$ &11.85$^{2.2}$ &4.22$^{1.43}$ &87.07$^{8.7}$ &77.44$^{9.1}$ &6.50$^{2.2}$ &1.57$^{0.5}$ &\textbf{23.55M}\\
		\textbf{C-Rend} &\textbf{83.49}$^{7.9}$ &\textbf{72.02}$^{8.8}$ &\textbf{11.70$^{3.0}$} &\textbf{3.99$^{1.1}$} &\textbf{87.78$^{10.4}$} &\textbf{78.64$^{9.8}$} &\textbf{6.45$^{1.9}$} &\textbf{1.23$^{0.8}$} &\textbf{23.55M}\\
		\bottomrule
	\end{tabular}
	\label{eval_ave3}
\end{table}

\begin{table}
	\centering
	\caption{Evaluation about follicle detection and counting}
	\begin{tabular}{c|ccc}
		\toprule
		\textbf{Method} & \textbf{FD($<$5mm)} &\textbf{MD($<$5mm)} &\textbf{Counting Error}\\
		\cline{1-4}
		Unet \cite{ronneberger2015u} &2.47$\pm9.7$ &16.07$\pm11.2$ &2.52$\pm3.8$\\
		Context\cite{tu2009auto} &3.68$\pm7.1$ &17.42$\pm12.1$ &2.27$\pm3.4$\\
		Edge\cite{chen2016dcan} &5.19$\pm5.2$ &17.69$\pm12.3$ &2.47$\pm2.9$\\
		Deeplab\cite{chen2017deeplab} &4.88$\pm9.6$ &18.60$\pm10.6$ &2.75$\pm2.7$\\
		Entropy\cite{wang2019boundary} &5.58$\pm8.7$ &19.60$\pm14.9$ &2.16$\pm3.0$\\
		\cline{1-4}	
		\textbf{Rend(ours)} &4.14$\pm7.4$ &13.78$\pm7.1$ &2.51$\pm2.1$\\
		\textbf{C-Rend(ours)} &\textbf{1.71$\pm5.4$} &\textbf{13.62$\pm8.6$} &\textbf{2.10$\pm1.9$}\\
		\bottomrule
	\end{tabular}
	\label{tab:FDMD_avg}
\end{table}
In Table \ref{tab:FDMD_avg}, we analyze the performance of different methods in detecting follicles with radius$<5mm$ and counting all follicles. We can observe that, embedding the rendering task and contrastive learning, C-Rend significantly reduces the FD (30\%), MD (15\%) and counting error (16\%) when compared to the baseline. C-Rend also surpasses other competitors in all criteria. Addressing the boundary ambiguity in C-Rend from point level may profoundly contribute to the localization of small follicles.

\begin{figure}
	\centering
	\includegraphics[width=\textwidth]{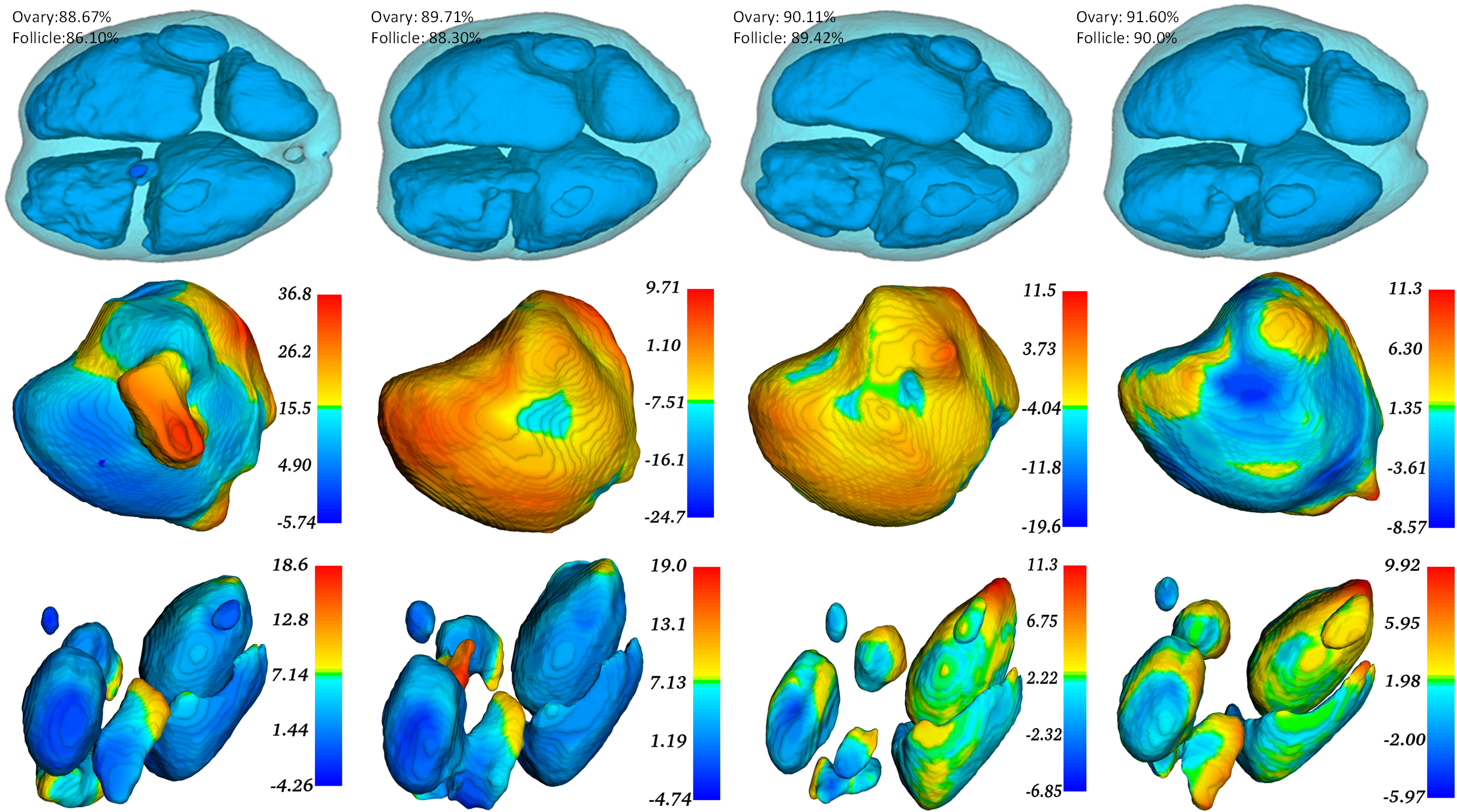}
	\caption{Visualization results. From left to right: U-net, \textit{Edge}, Rend and C-Rend. From Top to bottom: surface rendering of ovary (cyan) and follicles (blue), HD from ovary segmentation to ground truth, Hausdorff distance from follicle segmentation to ground truth. The colorbar is annotated with mean in the center, min and max on the ends.}
	\label{fig:3D-result}
\end{figure}

Fig. \ref{fig:3D-result} visualize the segmentation differences among the baseline U-net, \textit{Edge}, Rend and C-Rend. In surface rendering, C-Rend gets more clear boundary prediction than other methods for follicles with different scales. In HD surface distance, Rend and C-Rend present much lower mean/max/min distance errors for both ovary and follicles than the baseline and \textit{Edge} methods.

\section{Conclusion}
We proposed a general and lightweight framework for ultrasound images segmentation, which holds potentials for different deep architectures and applications. Exploiting point-level coarse prediction and fine-grained features, coupled with contrastive learning, to calibrate the ambiguous prediction on boundary are the main highlights of this work. The proposed C-Rend is extensively validated on the tough task of 3D ovarian US segmentation. In terms of efficacy and computational cost, both quantitative and qualitative evaluations support that, our C-Rend has clear advantages over other strong competitors in combating ambiguous boundary.

\subsubsection{Acknowledgments:}
This work was supported by the grant from National Key R\&D Program of China (No. 2019YFC0118300); Shenzhen Peacock Plan (No. KQTD2016053112051497, KQJSCX20180328095606003); Medical Scientific Research Foundation of Guangdong Province, China (No. B2018031); National Natural Science Foundation of China (No. NSFC61771130).


\bibliographystyle{splncs04}
\bibliography{paper}

\end{document}